\documentclass[journal=jctcce, manuscript=article, layout=twocolumn]{achemso}

\usepackage{txfonts}
\usepackage{color}
\usepackage{braket}
\definecolor{myblue}{rgb}{0,0,1}
\usepackage[breaklinks=true,colorlinks=true,linkcolor=myblue,urlcolor=myblue,citecolor=myblue]{hyperref}

\makeatletter
\let\l@addto@macro\relax
\makeatother
\usepackage[fontsize=11pt]{scrextend}

\usepackage{lineno}

\title{Focused Sampling for Low-Cost and Accurate\\ Ehrenfest Modeling of Cavity Quantum Electrodynamics}

\author{Ming-Hsiu Hsieh}
\author{Alex Krotz}
\author{Roel Tempelaar}
\email{roel.tempelaar@northwestern.edu}

\affiliation{Department of Chemistry, Northwestern University, 2145 Sheridan Road, Evanston, Illinois 60208, USA}

\begin{document}

\begin{tocentry}
\includegraphics{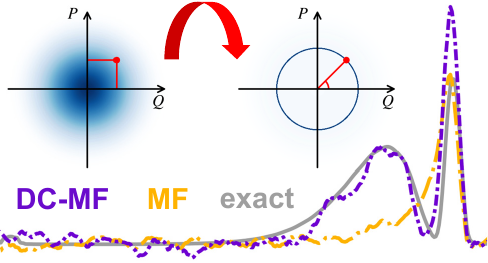}
\end{tocentry}

\begin{abstract}
An economic modeling approach for cavity quantum electrodynamics is provided by mean-field dynamics, wherein the optical field is described classically while a self-consistent interaction with quantum emitters is incorporated through the Ehrenfest theorem. However, conventional implementations of mean-field dynamics are known to suffer from a catastrophic leakage of zero-point energy, to lose accuracy in the short-cavity limit, and to require large numbers of trajectories to be sampled. Here, we address these three shortcomings within a single integrated approach. This approach builds on our recently-proposed modification of the Ehrenfest theorem, referred to as decoupled mean-field (DC-MF) dynamics, in combination with a focused sampling scheme that enforces zero-point energy at the single-trajectory level. The approach is shown to yield high accuracy in both short and long-cavity limits while reaching convergence within a minimal amount of trajectories.
\end{abstract}

\maketitle

\section{Introduction}

Cavity quantum electrodynamics (cQED) has taken center stage with the recent surge in studies showing cavity control of matter-based phenomena, including chemical reactions \cite{hutchisonModifyingChemicalLandscapes2012, thomasGroundStateChemicalReactivity2016, ribeiroPolaritonChemistryControlling2018, hertzogStrongLightMatter2019, hiraiRecentProgress2020, dunkelbergerVibrationCavityPolaritonChemistry2022, mandalTheoreticalAdvances2023} and energy transfer \cite{colesPolaritonmediatedEnergyTransfer2014, zhongNonRadiative2016, rozenmanLongRangeTransport2018, georgiouUltralongRange2021, pandyaTuningCoherentPropagation2022}. Early efforts to rationalize such cavity control resorted to minimal quantum representations, such as the Jaynes--Cummings \cite{jaynesComparisonQuantumSemiclassical1963} and Tavis--Cummings models \cite{tavisExactSolutionMolecule1968}, but it has since become evident that more elaborate multi-modal representations are necessary \cite{hoffmannEffectManyModes2020, wangRoadmapTheoryVibrational2021, tichauerMultiScale2021, balasubrahmaniyamCouplingDecouplingPolaritonic2021, fregoniTheoreticalChallengesPolaritonic2022, ribeiroMultimodePolaritonEffects2022, sanchez-barquillaTheoreticalPerspectiveMolecular2022, mandalMicroscopicTheoryMultimode2023, liMesoscaleMolecularSimulations2024,sunExploringDelocalizationDark2025} to capture collective effects, dark states, disorder, and dissipation \cite{ribeiroPolaritonChemistryControlling2018, scholesEmergence2021, engelhardtUnusualDynamical2022, changDarkState2024}. In such cases, retaining a full quantum-mechanical description of light and matter becomes prohibitively expensive, prompting the need to take the classical approximation for part of the involved coordinates.

A natural approach to model cQED under a partial classical approximation is by invoking classical electromagnetism, while retaining a quantum description of matter. In principle, this approach unifies established techniques from classical optics, including finite-difference time-domain \cite{yeeNumericalSolutionInitial1966} and finite element methods \cite{hrennikoffSolutionProblemsElasticity1941, courantVariationalMethodsSolution1943}, with those from quantum chemistry. Such unification can be realized through various nonadiabatic dynamics approaches based on quasiclassical \cite{liQuasiclassicalModelingCavity2020, SallerBenchmarkingQuasiclassicalMapping2021, heCommutatorMatrix2021} or mixed quantum--classical dynamics \cite{hoffmannCapturingVacuumFluctuations2019, hoffmannBenchmarkingSemiclassicalPerturbative2019}. Within the latter category, a straightforward means to ensure self-consistency of the light--matter interactions is provided by the Ehrenfest theorem, yielding a mean-field (MF) approach to cQED \cite{hoffmannCapturingVacuumFluctuations2019, hoffmannBenchmarkingSemiclassicalPerturbative2019}.

While MF dynamics is able to broadly capture trends arising from quantum emitters interacting with cavity modes \cite{hoffmannCapturingVacuumFluctuations2019, hoffmannBenchmarkingSemiclassicalPerturbative2019}, we have previously shown such interactions to lead to a catastrophic dissipation of cavity vacuum fluctuations under the application of the Ehrenfest theorem, yielding qualitatively-incorrect dynamics \cite{hsiehMeanFieldTreatmentVacuum2023, hsiehEhrenfestModelingCavity2023}. We have resolved this issue by modifying the Ehrenfest theorem such that vacuum fluctuations are selectively decoupled from the lowest-energy quantum state for all optical transitions of the quantum emitters. Referred to as decoupled mean-field (DC-MF) dynamics, this approach was shown to reach near-quantitative accuracy for examples involving two and three-level atoms embedded in a long cavity \cite{hsiehMeanFieldTreatmentVacuum2023, hsiehEhrenfestModelingCavity2023}. However, this left the short-cavity limit unaccounted for, where mixed quantum--classical dynamics may generally be expected to deliver a poor performance. Very recently, while investigating this limit, we discovered \cite{hsiehMixedQuantumClassicalDynamics2025} that MF dynamics captures the short-cavity limit remarkably well when the commonly-applied Wigner sampling of the initial classical cavity coordinates \cite{wignerOn1932} is replaced by a focused sampling scheme \cite{stockFlow1999, bonellaSemiclassicalImplementationMapping2003, dukeSimulatingExcitedState2015} that (approximately) enforces zero-point energy (ZPE) at the single-trajectory level. As an additional benefit, we found this approach to reduce the number of trajectories needed to reach converged results from thousands down to as little as one \cite{hsiehMixedQuantumClassicalDynamics2025}.

The abovementioned advances leave us with two distinct approaches reaching high accuracies in the short and long-cavity regimes, respectively, but there remains the need for a single low-cost methodology that reaches desirable accuracy for both short and long cavities. In this Article, we deliver such a methodology by incorporating focused sampling into DC-MF dynamics. We show such an implementation to necessitate a renormalization of matter-based transition dipole moments (TDMs) in the quantum--classical interaction Hamiltonian in order for it to reproduce the frequency of recurrent light--matter interactions in cQED. The resulting approach presents itself as a methodology that optimally balances computational efficiency and accuracy across cavity lengths.

\section{Theoretical Background}

We will first proceed to briefly summarize MF and DC-MF dynamics, as well as the employed model system, referring to our previous work for details \cite{hsiehMeanFieldTreatmentVacuum2023, hsiehEhrenfestModelingCavity2023}. We consider an atom interacting with the field of an optical cavity, as governed by the total mixed quantum--classical Hamiltonian
\begin{equation}
    \hat{H} = \hat{H}_\mathrm{A} + \hat{H}_\mathrm{AF} + H_\mathrm{F}. \label{eq:totalH}
\end{equation}
Here, the atomic Hamiltonian operator is given by
\begin{equation}
    \hat{H}_{\mathrm{{A}}} = \sum_{k}\epsilon_{k} \ket{k}\bra{k} ,\label{eq:HA_1}
\end{equation}
where $k$ labels the atomic energy levels, and $\epsilon_{k}$ represents the energy of level $k$. The atomic coordinates are represented by a quantum wavefunction, whose evolution is governed by the time-dependent Schr\"odinger equation,
\begin{equation}
    \ket{\dot{\psi}} = -\frac{i}{\hbar} \hat{H}\ket{\psi}. \label{eq:tdseMF}
\end{equation}

Within MF dynamics, the optical field of the cavity is described by the classical Hamiltonian function
\begin{equation}
    H_{\mathrm{F}} = \frac{1}{2}\sum_\alpha\left(P_\alpha^2+\omega_\alpha^2Q_\alpha^2\right),
\end{equation}
where $\alpha$ labels the cavity modes, and $\omega_\alpha$ denotes the corresponding angular frequency. Further, $Q_\alpha$ and $P_\alpha$ denote the corresponding position and momentum coordinates, which are associated with the electric and magnetic field components, respectively \cite{pellegriniOptimizedEffectivePotential2015}. As before \cite{flickAtomsMoleculesCavities2017, hoffmannCapturingVacuumFluctuations2019, hsiehMeanFieldTreatmentVacuum2023}, we consider the cavity to have a reduced dimensionality, invoking one light-propagation direction and one orthogonal polarization direction. Consequently, modes organize in a nondegenerate series, with the corresponding angular frequencies given by $\omega_{\alpha} = \pi c_0 \alpha / L$. Here, $c_0$ is the speed of light and $L$ is the cavity length. The atom--field interaction Hamiltonian operator is then given by
\begin{equation}
    \hat{H}_{\mathrm{AF}} = \sum_{\alpha}\sum_{k<l}\omega_{\alpha}\lambda_{\alpha}\mu_{kl}Q_{\alpha} \ket{k}\bra{l} + \mathrm{H.c.}, \label{eq:Haf}
\end{equation}
where $\mu_{kl}$ is the TDM between atomic levels $k$ and $l$, and H.c.~is short for Hermitian conjugate. Furthermore,
\begin{equation}
    \lambda_{\alpha} = \sqrt{\frac{2}{\epsilon_0L}}\sin\left(\frac{\pi \alpha r_\mathrm{A}}{L}\right)
\end{equation}
is the field amplitude of mode $\alpha$ at the atomic location $r_\mathrm{A}$ along the light-propagation direction, with $\epsilon_0$ as the vacuum permittivity. Here and henceforth, we assume the atom to reside at the center of the cavity, as a result of which this amplitude reduces to $\lambda_{\alpha}=\pm(2/\epsilon_0L)^{1/2}$ and $\lambda_{\alpha}=0$ for odd and even $\alpha$, respectively. The cavity coordinates are propagated using the Hamilton equations, with the force terms derived from the total Hamiltonian as
\begin{equation}
    F_{\alpha} = -\Braket{\psi|\frac{\partial \hat{H}}{\partial Q_{\alpha}}|\psi}.
\end{equation}

Whereas in MF dynamics the classical coordinates are representative of both vacuum fluctuations and thermal fluctuations, DC-MF dynamics separates those fluctuations into two separate coordinates. Accordingly, the optical field Hamiltonian function is expressed as \cite{hsiehMeanFieldTreatmentVacuum2023}
\begin{equation}
    H^{\mathrm{DC}}_\mathrm{F} = \frac{1}{2}\sum_{\alpha}\left(p_{\alpha}^{2}+\omega_{\alpha}^{2}q_{\alpha}^{2}+P_{\alpha}^{2}+\omega_{\alpha}^{2}Q_{\alpha}^{2}\right),
\end{equation}
with the thermal and vacuum positions denoted as $q_\alpha$ and $Q_\alpha$, respectively, and analogously for the momenta $p_\alpha$ and $P_\alpha$. (We adopted a change in notation compared to Refs.~\citenum{hsiehMeanFieldTreatmentVacuum2023} and \citenum{hsiehEhrenfestModelingCavity2023}, where the vacuum and thermal coordinates were expressed as upper case characters with and without tilde, respectively. This notation change simplifies the discussion in the present Article.) These coordinates are separately evolved through the Hamilton equations, with force terms given by
\begin{equation}
    f_{\alpha} = -\Braket{\psi|\frac{\partial \hat{H}}{\partial q_{\alpha}}|\psi}, \qquad
    F_{\alpha} = -\Braket{\psi|\frac{\partial \hat{H}}{\partial Q_{\alpha}}|\psi}.
\end{equation}
For every optical transition, the vacuum positions $Q_\alpha$ are then decoupled from the lower level of the atom, yielding an atom--field interaction Hamiltonian of the form \cite{hsiehEhrenfestModelingCavity2023}
\begin{equation}
    \hat{H}^{\mathrm{DC}}_{\mathrm{AF}} = \sum_{\alpha}\sum_{k<l}\omega_{\alpha}\lambda_{\alpha}\mu_{kl}\left(q_{\alpha}+\rho_l Q_{\alpha}\right)\ket{k}\bra{l} + \mathrm{H.c.} \label{eq:new_Haf}
\end{equation}
Here, the term $\rho_l \equiv \braket{\psi|l}\braket{l|\psi}$ represents the occupancy of atomic level $l$. It is responsible for a gradual decoupling of vacuum contributions to the lower level in each transition. This decoupling becomes absolute once the atom is in the lowest state associated with a given transition.

\begin{figure}[b!]
  \includegraphics{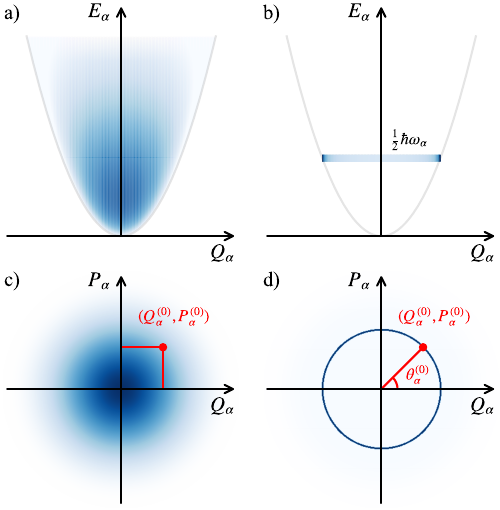}
  \caption{Schematic illustration of the probability distributions for a single cavity mode (denoted $\alpha$) under Wigner sampling (a,c) and under focused sampling (b,d). Shown as a heat map are the distributions as a function of position $Q_\alpha$ and mode energy $E_\alpha = \frac{1}{2}(P_\alpha^2 + \omega_\alpha^2 Q_\alpha^2)$ (a,b) and as a function of position $Q_\alpha$ and momentum $P_\alpha$ (c,d).}
  \label{fig_Wig_focused}
\end{figure}

As before \cite{flickAtomsMoleculesCavities2017, hoffmannCapturingVacuumFluctuations2019, hsiehMeanFieldTreatmentVacuum2023}, we consider the zero-temperature limit, such that thermal fluctuations of the optical field are omitted when initializing the atom--cavity system and only the vacuum fluctuations need to be accounted for. Arguably, the most common approach to initialize classical coordinates representative of vacuum fluctuations is by means of Wigner sampling, which is schematically illustrated for a single cavity mode in Fig.~\ref{fig_Wig_focused} (a,c). Denoting initial values of classical coordinates with superscript $(0)$ as a short-hand notation for $t=0$, the initial position and momentum coordinates of the mode are stochastically drawn from the normal distribution
\begin{equation}
    P({Q_\alpha^{(0)}, P_\alpha^{(0)}}) = \frac{1}{\pi\hbar}\mathrm{exp}\left( -\frac{{P^{(0)}_\alpha}^2}{\hbar\omega_\alpha} - \frac{\omega_\alpha {Q^{(0)}_\alpha}^2}{\hbar}\right).
    \label{eq:zero_Wig}
\end{equation}
Under such sampling, the ensemble average reproduces the ground-state quantum harmonic oscillator wavefunction, as stipulated by the Wigner quasiprobability distribution \cite{wignerOn1932}. While ZPE is thus reproduced at the ensemble level, single trajectories will initialize that undershoot or overshoot this energy.

\section{Short-Cavity Regime}

\begin{figure}[t!]
  \includegraphics{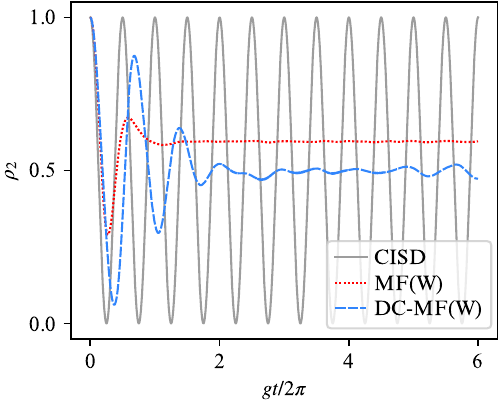}
  \caption{Transient excited-state occupancy of a two-level atom embedded in a half-wavelength cavity, resulting from MF(W) (red) and DC-MF(W) (blue) dynamics. Results from CISD are also shown (gray).}
  \label{fig_Rho_small}
\end{figure}

Recently, we found MF dynamics under Wigner sampling to perform poorly in the short-cavity limit \cite{hsiehMixedQuantumClassicalDynamics2025}. Specifically, we considered a two-level atom embedded in a cavity whose length was taken to amount to half the wavelength associated with the atom's optical transition, so that the fundamental mode of the cavity has an angular frequency satisfying
$\omega_1=(\epsilon_{2}-\epsilon_{1})\hbar^{-1}$. Under such resonant conditions, the atom interacts significantly only with this fundamental mode (so long as the ultra or deep-strong coupling regime is avoided). Absent any dissipation, the transient atomic (and optical) occupancies are then expected to exhibit a persistent Rabi oscillation at a frequency $2g$, where we define 
\begin{equation}
    g\equiv \mu_{12}\lambda_1\sqrt{\omega_1/2\hbar},
    \label{eq:g_definition}
\end{equation}
with $\lambda_1$ as the field amplitude of the fundamental mode. However, instead of such a persistent Rabi oscillation, MF dynamics under Wigner sampling showed a marked decay \cite{hsiehMixedQuantumClassicalDynamics2025}. We proceed by revisiting this simulation, while referring to the applied methodology as MF(W) dynamics.

The result is reproduced in Fig.~\ref{fig_Rho_small}. Here, a comparison is drawn of the transient atomic excited-state occupancy, $\rho_2$, with full-quantum results produced by configuration interaction singles and doubles (CISD), details of which can be found in Ref.~\citenum{hsiehMeanFieldTreatmentVacuum2023} and the Supporting Information (SI). The parameters are chosen such that $\omega_1 = 50 g$, which falls within the regime where the rotating wave approximation is applicable. Since $\omega_1=(\epsilon_{2}-\epsilon_{1})\hbar^{-1}$, and with the transient occupancy depicted as a function of $gt$, the result as shown does not depend on the numerical values taken for the parameters. An average of a total of 50\;000 trajectories was taken in order to ensure convergence. Beyond revisiting these previously-reported results, we also include in Fig.~\ref{fig_Rho_small} the outcome of DC-MF dynamics where vacuum coordinates are sampled from a Wigner distribution while thermal coordinates are initialized as zero, which we refer to as DC-MF(W) dynamics. Here, an averaging is taken over a total of 10\;000 trajectories. Whereas the oscillation decay is seen to be somewhat less severe compared to MF(W) dynamics, the overall performance is still poor, with a conspicuous underestimation of the overall Rabi oscillation frequency.

Previously \cite{hsiehMixedQuantumClassicalDynamics2025}, we were able to attribute the Rabi oscillation decay observed for MF(W) dynamics to a destructive interference arising from a dependence of the oscillation frequency on the initial energy of the cavity mode. This prompted us to replace the Wigner sampling scheme with one where the mode energy is fixed to the value that reproduces the expected Rabi frequency. We refer to such a enforcement of properties at the single-trajectory level as focused sampling, drawing from analogous terminology coined in previous works on photo-induced chemical processes \cite{bonellaSemiclassicalImplementationMapping2003, dukeSimulatingExcitedState2015}. In case of the latter, sampling was focused so as to ensure that each trajectory was initialized on a single electronic surface \cite{bonellaSemiclassicalImplementationMapping2003, dukeSimulatingExcitedState2015}. Instead, our applied focusing scheme restricts the cavity mode to a given initial energy. This in turn draws inspiration from previous quasiclassical studies, where restricting modes to the ZPE was found to yield improved accuracy \cite{schatzOriginCrossSection1983, luClassicalMechanicsIntramolecular1989, ben-nunZeroPointEnergy1996, stockFlow1999}.

Notably, while we resorted to focused sampling primarily in order to resolve unphysical behavior arising from destructive interference under Wigner sampling, in many instances such interference leaves observables unchanged. It does, however, yield an unnecessary increase in the computational cost of a simulation. As a result, focused sampling techniques are commonly employed for the sake of computational cost reduction \cite{dukeSimulatingExcitedState2015, mandalQuasiDiabatic2018, runesonSpinMapping2019, ananthPathIntegrals2022}, with reported improvement factors lying between 10 (Ref.~\citenum{kimQuantumClassical2008}) and 25 (Ref.~\citenum{bonellaLinearizedPath202005}). 

Our employed focused sampling scheme is schematically illustrated for a single cavity mode in Fig.~\ref{fig_Wig_focused} (b,d). Accordingly, initial values of the classical coordinates are drawn from the phase-space circle associated with a given energy through sampling of an angle $\theta_\alpha^{(0)}$, as
\begin{eqnarray}
    \theta_\alpha^{(0)} &\in& [0, 2\pi), \nonumber\\
    Q_\alpha^{(0)} &=& \sqrt{\frac{2n^{(0)} \hbar}{\omega_\alpha}} \cos\theta_\alpha^{(0)}, \label{eq:focusedsampling} \\
    P_\alpha^{(0)} &=& \sqrt{2n^{(0)} \hbar \omega_\alpha} \sin\theta_\alpha^{(0)}, \nonumber
\end{eqnarray}
with $n^{(0)}$ as the initial (classical) occupancy of the cavity mode. Denoting the initial energy contained in this mode as $E_\alpha^{(0)}$, this occupancy equals $E^{(0)}_\alpha/\hbar\omega_\alpha$. Upon fixing $n^{(0)}=1/2$, the initial mode energy thus equals $\hbar\omega_\alpha/2$, meaning that ZPE is enforced exactly.

\begin{figure*}
  \includegraphics{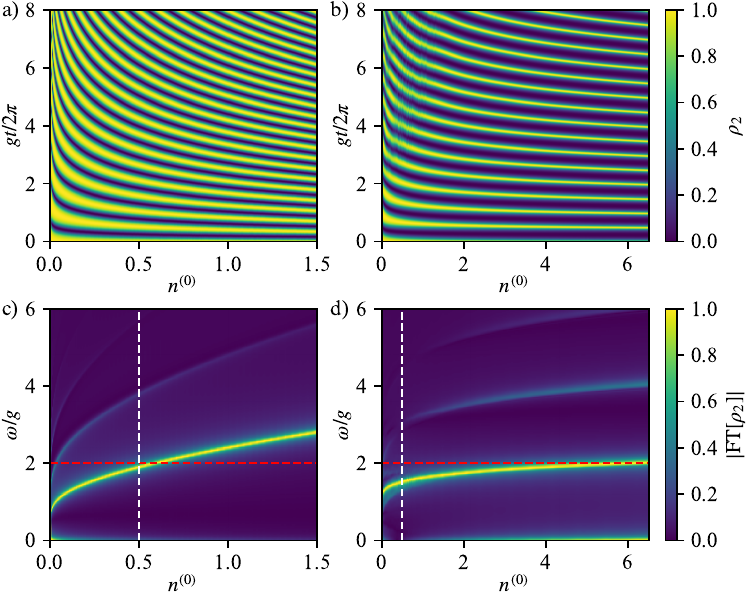}
  \caption{Transient excited-state occupancy of the two-level atom embedded in a half-wavelength cavity resulting from MF dynamics (a) and DC-MF dynamics (b) as a function of the initial optical occupancy, $n^{(0)}$. Also shown are the corresponding Fourier amplitudes (c,d). Indicated with dashes are the expected Rabi oscillation frequency at $\omega = 2g$ (red) and the optical occupancy corresponding to the ZPE at $n^{(0)}=1/2$ (white).
  }
  \label{fig_Rho_N0scan}
\end{figure*}

With $\omega_\alpha\gg g$, the cavity mode browses the full range of angles well before it begins to exchange energy with the atom. As a result, the atom--field dynamics is independent of $\theta_\alpha^{(0)}$, and is dependent only on the initial optical occupancy, $n^{(0)}$. In Fig.~\ref{fig_Rho_N0scan} (a), we characterize this dependence by showing the transient atomic excited-state occupancy, $\rho_2$, as a function of $n^{(0)}$, as calculated within MF dynamics. Previously \cite{hsiehMixedQuantumClassicalDynamics2025}, we reported a similar result, but there the MF dynamics was solved semi-analytically within the rotating wave approximation. Here, we instead conduct a fully-numerical simulation to ensure consistency with the other results presented in this Article (for which a semi-analytical treatment is not feasible), where it should be noted that the MF dynamics result shown in Fig.~\ref{fig_Rho_N0scan} (a) is indistinguishable from our previous treatment. We can thus resort to the same arguments as made previously \cite{hsiehMixedQuantumClassicalDynamics2025} in order to rationalize the observed trends.

Depicted in Fig.~\ref{fig_Rho_N0scan} (c) is the Fourier amplitude of the transient occupancy $\rho_2$. The underlying Fourier transform was performed over a duration of $gt = 200$, after multiplication of $\rho_2$ by an exponential with an $1/e$ time of $gt = 20$ to avoid clipping and after subtracting a 0.5 baseline, upon which the signal was normalized. The Fourier amplitude shows a dominant component whose energy increases with $n^{(0)}$, alongside a pair of additional components that vanish with increasing $n^{(0)}$. All components are seen to reflect the aforementioned dependence of the Rabi oscillation frequency on the initial energy of the cavity mode, which ultimately causes destructive interference under Wigner sampling. As shown previously \cite{hsiehMixedQuantumClassicalDynamics2025}, the dominant component follows the frequency relationship $\omega\approx2g(n^{(0)}+1/2)^{1/2}$, but its frequency undershoots this equality in the limit $n^{(0)}\rightarrow 0$, in order to reach $\omega=0$ for $n^{(0)}=0$. The latter reflects that mixed quantum--classical dynamics uniquely predicts a complete absence of emission by the atom when the optical field is unoccupied \cite{liMixedQuantumclassicalElectrodynamics2018}. (We note that spontaneous emission is achieved within mixed quantum--classical dynamics for any $n^{(0)}>0$, and that this process does not rely on further quantization of the optical mode.) As a consequence, the dominant component crosses the ZPE at an occupancy slightly exceeding $n^{(0)}=1/2$. Previously \cite{hsiehMixedQuantumClassicalDynamics2025}, we therefore adopted an adjusted value of $n^{(0)}=0.59$. This value can be inferred from Fig.~\ref{fig_Rho_N0scan} (c) based on the intersection of the $n^{(0)}$-dependent dominant component with the expected Rabi frequency, $\omega =2g$.

A similar analysis is presented in Fig.~\ref{fig_Rho_N0scan} (b,d), but for DC-MF dynamics. Whereas the MF dynamics results were obtained using a single trajectory, this simulation required an averaging over a total of 10 trajectories in order to ensure convergence. Similarly to MF dynamics, anharmonicities are apparent, but otherwise there are a number of comparative differences seen in the results. First, the dominant component shows an energy increase with $n^{(0)}$ that is much slower compared to what was found for MF dynamics, yielding a ZPE crossing at $n^{(0)}\approx6$. Second, anharmonicities do not generally vanish with increasing $n^{(0)}$, but instead are organized in a variety of components, some vanishing and some growing in. Combined, however, these components leave a window around $n^{(0)}\approx 1/2$, where Rabi oscillations are reasonably harmonic. Altogether, these results show that having DC-MF dynamics reproduce Rabi oscillations by adjusting $n^{(0)}$ is not a feasible approach, owing to the large $n^{(0)}$ value necessary, and the strong anharmonicity present at this value.

An alternative and trivial means to affect the Rabi oscillation frequency is through a renormalization of the atomic TDM in the light--matter interaction Hamiltonian. Accordingly, we perform the replacement $\mu_{kl}\leftarrow\gamma\mu_{kl}$
in Eqs.~\ref{eq:Haf} and \ref{eq:new_Haf}, where $\gamma$ denotes a rescaling factor. This allows one to reproduce the expected Rabi frequency through variations of $\gamma$, while focusing the mode occupancy strictly to $n^{(0)}=1/2$. Not only is this value of $n^{(0)}$ physically-informed, as it enforces ZPE exactly for the mode, it also ensures consistency with analogous focusing approaches applied previously in quasiclassical modeling studies \cite{schatzOriginCrossSection1983, luClassicalMechanicsIntramolecular1989, ben-nunZeroPointEnergy1996, stockFlow1999}, where it has been argued that any trajectory with mode energies more or less than the ZPE is unphysical. Moreover, at this value the anharmonicities seen in Fig.~\ref{fig_Rho_N0scan} are minimal. The applicable rescaling factor $\gamma$ can be obtained by a similar intersection procedure applied to Fig.~\ref{fig_Rho_N0scan} (c,d), and is determined as the ratio of the expected Rabi frequency ($\omega=2g$), divided by the frequency where the dominant component crosses $n^{(0)}=1/2$. This yields $\gamma=1.06$ and $\gamma=1.33$ for MF and DC-MF dynamics, respectively. Conceptually, such a rescaling points to an effective underestimation of the dipolar effect under the applied mixed quantum--classical schemes. (Alternatively, one could associate this rescaling factor with the field amplitudes, $\lambda_\alpha$. However, the latter is related to the atomic location along the light-propagation direction, which we expect not to be affected by the partial classical approximation.)

\begin{figure}[t!]
  \includegraphics{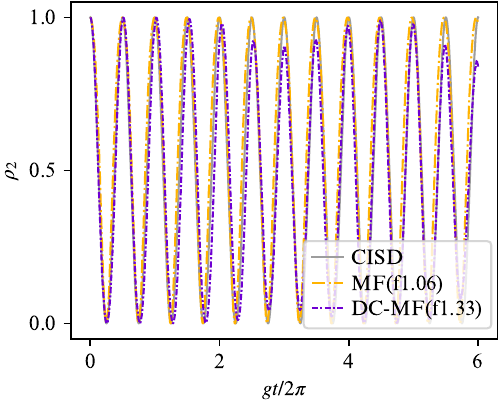}
  \caption{Same as Fig.~\ref{fig_Rho_small}, but for results from MF(f1.06) (orange) and DC-MF(f1.33) (purple) dynamics.}
  \label{fig_Rho_small_f}
\end{figure}

In the following, we will invoke this rescaling approach while applying focused sampling to MF and DC-MF dynamics, which henceforth will be referred to as MF(f1.06) and DC-MF(f1.33) dynamics, respectively. Our particular choice of rescaling factors is affirmed by Fig.~\ref{fig_Rho_small_f}, which presents results from MF(f1.06) and DC-MF(f1.33) dynamics showing favorable agreement with CISD results, although we note the minor beating pattern exhibited by DC-MF(f1.33) dynamics due to the presence of low-frequency anharmonicity. Evidently, focused sampling is seen to effectively resolve the spurious destructive interference caused by Wigner sampling for both methods, and the renormalization of TDMs in the light--matter interaction Hamiltonian helps reproduce the correct Rabi frequency. (Analogous results for MF and DC-MF dynamics under focused sampling but without TDM renormalizations are presented in Fig.~S1 of the SI, showing persistent Rabi oscillations, but at an underestimated frequency.)

\section{Long-Cavity Regime}

Having established that MF(1.06) and DC-MF(f1.33) dynamics yield desirable results in the short-cavity limit, we proceed to revisit the long-cavity regime for which DC-MF dynamics was originally developed \cite{hsiehMeanFieldTreatmentVacuum2023, hsiehEhrenfestModelingCavity2023}. Under Wigner sampling, DC-MF dynamics was previously shown to reach near-quantitative accuracy in this regime \cite{hsiehMeanFieldTreatmentVacuum2023, hsiehEhrenfestModelingCavity2023}, while MF dynamics proved broadly capable of reproducing qualitative trends \cite{hoffmannCapturingVacuumFluctuations2019, hoffmannBenchmarkingSemiclassicalPerturbative2019}. It would thus be interesting to make a comparative assessment of the performance of both methods under focused sampling and TDM renormalizations.

As in our previous evaluations of the long-cavity regime \cite{hsiehMeanFieldTreatmentVacuum2023, hsiehEhrenfestModelingCavity2023}, we parametrize our atom--cavity system following the example of Hoffmann \emph{et al.} \cite{hoffmannCapturingVacuumFluctuations2019}, such that the atom is taken to represent the lowest two levels of a one-dimensional hydrogen model \cite{suModelAtomMultiphoton1991, flickAtomsMoleculesCavities2017}, with energies $\epsilon_1= -0.6738$ a.u.~and $\epsilon_2=-0.2798$ a.u., and with a TDM $\mu_{12} = 1.034$ a.u., whereas the cavity length is taken as $L=2.362\times 10^{5}~\mathrm{a.u.}~=12.5~\mathrm{\mu m}$. The 400 lowest-energy cavity modes are incorporated in our simulations, which suffices to ensure convergence.

\begin{figure} [b!]
  \includegraphics{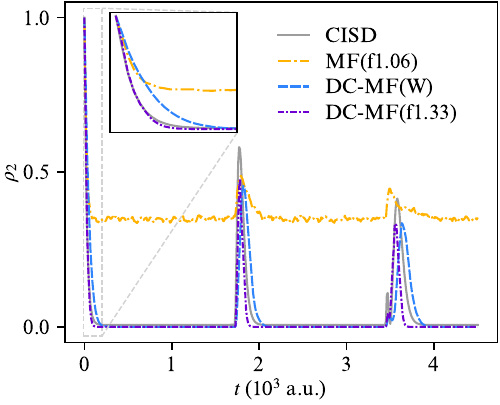}
  \caption{Transient excited-state occupancy of a two-level atom embedded in a long cavity ($L=12.5~\mathrm{\mu}\mathrm{m}$). Shown are results from MF(f1.06) (orange), DC-MF(W) (blue), and DC-MF(f1.33) (purple) dynamics, compared against those from CISD (gray). The inset shows early-time detail.}
  \label{fig_Rho_Nph}
\end{figure}

First, we report that focusing the initial cavity mode occupancies as $n^{(0)} = 1/2$ yields results virtually indistinguishable from those from Wigner sampling for both MF and DC-MF dynamics (see Fig.~S2 in the SI). This implies that the destructive interference seen for Rabi oscillations in the half-wavelength cavity does not negatively affect the long-cavity limit. We consider this a desirable outcome, since DC-MF(W) dynamics did already perform favorably for this long-cavity implementation \cite{hsiehMeanFieldTreatmentVacuum2023}, although one could have wished that the change in sampling approach would resolve inaccuracies seen for MF(W) dynamics \cite{hoffmannCapturingVacuumFluctuations2019, hsiehMeanFieldTreatmentVacuum2023, hsiehEhrenfestModelingCavity2023}. Regardless, such switching from Wigner sampling to focused sampling allows the number of trajectories, necessary for convergence, to be reduced from (roughly) 500 to 30 for DC-MF dynamics, while a reduction from 5000 to 2500 was found for MF dynamics. 

With regard to modeling accuracy, things turn interesting once we apply the aforementioned TDM renormalizations to both methodologies. Fig.~\ref{fig_Rho_Nph} compares DC-MF(f1.33) dynamics to DC-MF(W) dynamics where, remarkably, renormalizing the TDM is seen to almost fully remedy the lag known \cite{hsiehMeanFieldTreatmentVacuum2023, hsiehEhrenfestModelingCavity2023} to occur for the atomic excited-state occupancy within DC-MF(W) dynamics. As such, the applied renormalization yields a further-improved agreement of this method with CISD results, with minor discrepancies occurring exclusively in the reabsorption events. Also included in the comparison is MF(f1.06) dynamics, results of which show qualitative behavior consistent with previous MF(W) dynamics simulations but with a slightly improved relaxation rate \cite{hoffmannCapturingVacuumFluctuations2019, hoffmannBenchmarkingSemiclassicalPerturbative2019}.

\begin{figure}[t]
  \includegraphics{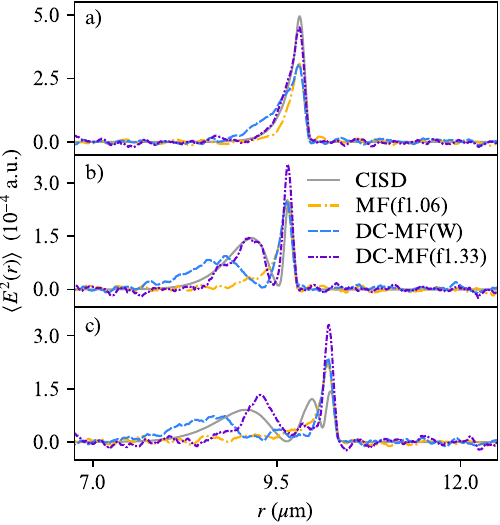}
  \caption{Optical field intensity at $t = 500$~a.u.~(a), $t = 2200$~a.u.~(b), and $t = 4000$~a.u.~(c), which correspond to instances shortly after the first emission, the first reabsorption, and the second reabsorption events, respectively. Note that the optical field is symmetric with respect to reflection through the cavity center, and only the right half of the field intensity is depicted. Shown are results from MF(f1.06) (orange), DC-MF(W) (blue), and DC-MF(f1.33) (purple) dynamics, compared against those from CISD (gray). Each trace was smoothed by a $0.125$ $\mathrm{\mu}$m moving average for ease of demonstration.}
  \label{fig_Int}
\end{figure}

Going beyond atomic occupancies, the spatially-resolved optical field intensity poses a more challenging test of the accuracy of dynamical methods for cQED (albeit one that is not always the most practically relevant). It is thus worth assessing the accuracy of MF(f1.06) and DC-MF(f1.33) dynamics in reproducing the optical field intensity, which is evaluated through the expression
\begin{eqnarray}
    \Braket{E^{2}(r)} &=&  \Braket{\Big(\sum_{\alpha}\sqrt{2\omega_{\alpha}\hbar^{-1}}\zeta_\alpha q_\alpha\Big)^2} \\ \nonumber&+&\Braket{\Big(\sum_{\alpha}\sqrt{2\omega_{\alpha}\hbar^{-1}}\zeta_\alpha Q_\alpha\Big)^2}- \sum_{\alpha} \zeta_\alpha^2 .
\end{eqnarray}
Here, the spatial mode function is given by
\begin{equation}
    \zeta_{\alpha}(r)=\sqrt{\frac{\hbar\omega_{\alpha}}{\epsilon_0L}}\sin\left(\frac{\pi \alpha}{L} r\right).
\end{equation}
Convergence of the optical field intensity with respect to the number of trajectories is much more demanding than that of atomic occupancies. Indeed, 500\;000 and 100\;000 trajectories are required for MF dynamics under Wigner and focused sampling, respectively, whereas 100\;000 and 50\;000 trajectories are required for DC-MF dynamics.

As with the atomic occupancies, optical field intensity profiles generated by MF dynamics and DC-MF dynamics are invariant with regard to whether Wigner sampling or focused sampling is applied (see Fig.~S3 in the SI), but renormalizing the TDM has a significant effect on DC-MF dynamics. Shown in Fig.~\ref{fig_Int} are optical field intensity results for MF(f1.06), DC-MF(W), and DC-MF(f1.33) dynamics, compared against CISD results at different time instances. While DC-MF(W) dynamics captures most qualitative features seen in the CISD results, as discussed before \cite{hoffmannCapturingVacuumFluctuations2019, hsiehMeanFieldTreatmentVacuum2023}, it is again remarkable to see the extent to which the TDM renormalization improves the accuracy of this method to a near-quantitative level. Indeed, DC-MF(f1.33) dynamics reproduces subtle details of the lagging wavefronts that arise upon reabsorption \cite{hsiehMeanFieldTreatmentVacuum2023}, with differences creeping in only after two reabsorption events. Combined with the favorable results obtained for the half-wavelength cavity, DC-MF(f1.33) dynamics thus presents itself as a method with applicability across cavity lengths. Moreover, the identical values of the applicable TDM rescaling factor $\gamma$ between the short-cavity and long-cavity regimes offer an optimistic prospect regarding the generality of this factor across diverse problems.

\section{Conclusions and Outlook}

In conclusion, DC-MF dynamics offers a route to accurate and low-cost simulations of cQED across cavity lengths, provided that focused sampling and TDM renormalizations are invoked. In demonstrating this, we built heavily on recent work \cite{hsiehMixedQuantumClassicalDynamics2025} where MF dynamics (without the decoupling scheme) was shown to reproduce full-quantum results for a half-wavelength cavity. Such was achieved under a focused sampling of initial optical field coordinates, in combination with a slight renormalization of the initial optical ZPE. Here, we present a renormalization of the TDMs as an alternative renormalization procedure that not only produces equally accurate results for MF dynamics, but that also allows DC-MF dynamics to produce desirable results. As mentioned, we find that switching from the commonly-used Wigner sampling technique to focused sampling has only a minor effect on Ehrenfest-based mixed quantum--classical modeling in the long-cavity regime, meaning that it fortuitously does not compromise the high accuracy that DC-MF dynamics is known to reach in this regime \cite{hsiehMeanFieldTreatmentVacuum2023, hsiehEhrenfestModelingCavity2023}. TDM renormalizations, on the other hand, provide a further source of accuracy improvement for this method, yielding atomic occupancies and optical field intensity profiles in excellent agreement with full-quantum results.

We consistently find focused sampling to dramatically improve the accuracy of both MF and DC-MF dynamics in the short-cavity regime, yet previous applications of this sampling technique were instead primarily concerned with a reduction of computational cost \cite{stockFlow1999, bonellaSemiclassicalImplementationMapping2003, bonellaLinearizedPath202005, kimQuantumClassical2008, mandalQuasiDiabatic2018, runesonSpinMapping2019, ananthPathIntegrals2022}. While we did not systematically study the computational cost savings afforded by focused sampling in cQED, which we reserve for a future study, we anecdotally found such savings to oftentimes be significant, especially in the short-cavity regime, where a reduction of the necessary number of trajectories was found of three orders of magnitude. In the long-cavity regime, this reduction was much more modest, with a factor of two being typical. Moreover, we find the number of trajectories necessary for convergence to be strongly dependent on the observable under consideration. Notably, transient occupancies of quantum emitters require about 10 to 30 trajectories for DC-MF dynamics under focused sampling, a number much lower than that typically used in mixed quantum--classical simulations of population dynamics. For the optical field intensity, this number raises to values in the range 50\;000 to 100\;000, which are more typical.

The results presented in this Article help to further establish DC-MF dynamics as a mixed quantum--classical approach for cQED simulations that optimally balances accuracy and computational efficiency. This not only promotes the use of DC-MF dynamics in studying cQED based on model systems, but it also bears relevance to a broader effort to incorporate quantum modeling in established techniques from classical optics \cite{mengFrequency2013, sakkoDynamical2014, ruggenthalerFrom2018, jestadtLight2019, sukharevEfficient2023, weightTheory2023}, including finite-difference time-domain, for which DC-MF dynamics offers a preferred route. The cost associated with trajectory sampling is very modest, so long as transient matter-based occupancies are being considered, while the evaluation of optical field intensity profiles remains computationally much more challenging.

Against the favorable performance of DC-MF dynamics, there are various aspects of this method that await a deeper rationale. One such aspect is concerned with the form of the decoupling term appearing in Eq.~\ref{eq:new_Haf}. Broadly speaking, this term may be considered representative of a functional of the atomic quantum state, which vanishes when the ground state is occupied while reaching unity for a fully-occupied excited state \cite{hsiehMeanFieldTreatmentVacuum2023}. Future efforts may unveil the existence of an exact functional, or at least identify improved proxies that further enhance the method's accuracy. Another aspect that deserves further attention is concerned with the TDM rescaling factors introduced in the present Article. Ultimately, these factors compensate for the tendency of mixed quantum--classical modeling to overshoot the optical occupancy needed to reproduce Rabi oscillations. This in turn is due to the oscillation frequency decaying to zero in the limit $n^{(0)}\rightarrow0$ \cite{hsiehMixedQuantumClassicalDynamics2025}. It would be interesting to further investigate this behavior, possibly through an analysis similar to that presented in Ref.~\citenum{hsiehMixedQuantumClassicalDynamics2025}. Altogether, such inquiries may provide further guidance for the application of mixed quantum--classical dynamics to the modeling of transient phenomena.

\section*{Supporting Information}
Methodological details of CISD dynamics and focused sampling results without TDM renormalizations.

\section*{Acknowledgement}
The authors thank Ken Miyazaki for fruitful discussions. This material is based upon work supported by the National Science Foundation under Grant No.~2145433. Research reported in this publication was supported, in part, by the International Institute for Nanotechnology at Northwestern University. M.-H.H.~gratefully acknowledges support from the Ryan Fellowship and the International Institute for Nanotechnology at Northwestern University.
 
\bibliography{bib}

\end{document}


\maketitle

\tableofcontents

\newpage

\section{Configuration Interaction Singles and Doubles}

The full-quantum results reported in the main text were obtained by using configuration interaction singles and doubles (CISD). Accordingly, the total Hamiltonian takes the form
    \begin{equation}
    \hat{H} = \hat{H}_\mathrm{A} + \hat{H}_\mathrm{AF} + \hat{H}_\mathrm{F}, \label{eq:totalH}
    \end{equation}
with $\hat{H}_\mathrm{A}$ defined by Eq.~2 of the main text, and with the atom--field interaction Hamiltonian given by
    \begin{equation}
    \hat{H}_{\mathrm{AF}} = \sum_{\alpha}\sum_{k<l}\omega_{\alpha}\lambda_{\alpha}\mu_{kl}\hat{Q}_{\alpha} \ket{k}\bra{l} + \mathrm{H.c.} \label{eq:Haf}
    \end{equation}
Here, $\hat{Q}_\alpha=\sqrt{\hbar/2\omega_\alpha}\left(\hat{a}_\alpha^\dagger + \hat{a}_\alpha\right)$ is the position-like operator of cavity mode $\alpha$, expressed in terms of the associated creation and annihilation operators. Adopting such ladder operators, the optical field Hamiltonian is expressed as 
    \begin{equation}
    \hat{H}_{\mathrm{F}} =  \sum_\alpha\hbar\omega_\alpha\left(\hat{a}^\dagger_\alpha \hat{a}_\alpha + \frac{1}{2}\right).
    \end{equation}

Within the applied CISD formalism, the tensor products of the optical field states are truncated to a total of two quanta (while retaining the full set of atomic states). Accordingly, the total wavefunction is expanded as
    \begin{equation}
    \ket{\psi} = \sum_{k} c_{k,\varnothing} \ket{k} \otimes \ket{\varnothing} + \sum_{k}\sum_{\alpha} c_{k,\alpha} \ket{k} \otimes \hat{a}^{\dagger}_\alpha \ket{\varnothing} + \sum_{k}\sum_{\alpha,\beta} c_{k,\alpha,\beta} \ket{k} \otimes \hat{a}^{\dagger}_\alpha \hat{a}^{\dagger}_\beta \ket{\varnothing},
    \end{equation} 
where $\ket{\varnothing}$ denotes the vacuum state of the optical field. The wavefunction is initialized in the atomic excited state as $\ket{\psi} = N\ket{2} \braket{1|G}$, where $G$ denotes the lowest eigenstate of $\hat{H}$, and $N$ is a normalization constant. The wavefunction is then propagated according to the time-dependent Schr\"odinger equation as Eq.~3 of the main text. The atomic excited-state population is calculated by $\rho_2 = \braket{\psi|2}\braket{2|\psi}$. Unless notes otherwise, the parameters employed in the exact quantum reference calculations are the same as those applied in MF and DC-MF dynamics.


\section{Focused Sampling Results without TDM Renormalizations}

\begin{figure}[!h]
\includegraphics{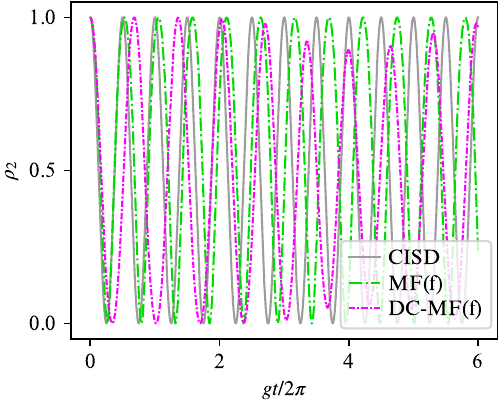}
  \caption{Same as Fig.~4 of the main text, but for MF(f) (green) and DC-MF(f) (magenta) dynamics.
  }
  \label{small}
\end{figure}

\begin{figure}[!h]
\includegraphics{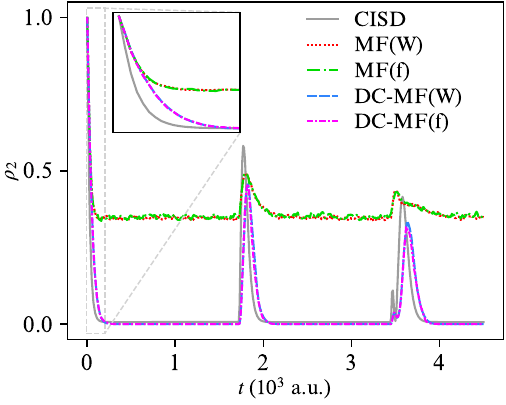}
  \caption{Same as Fig.~5 of the main text, but showing results from MF(W) (red), MF(f) (green), DC-MF(W) (blue), and DC-MF(f) (magenta) dynamics.
  }
  \label{long_population}
\end{figure}

\begin{figure}[!h]
\includegraphics{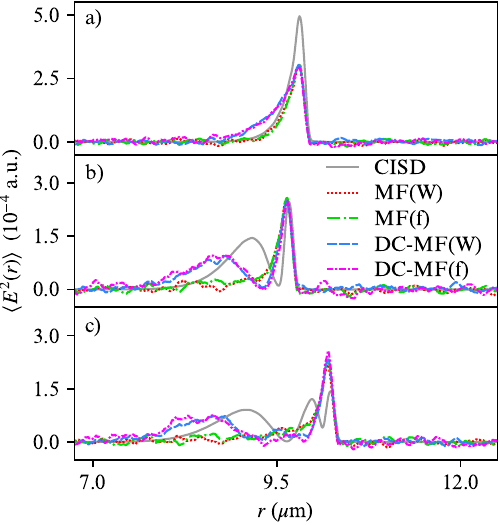}
  \caption{Same as Fig.~6 of the main text, but showing results from MF(W) (red), MF(f) (green), DC-MF(W) (blue), and DC-MF(f) (magenta) dynamics.
  }
  \label{long_intensity}
\end{figure}

Here, we complement the results shown in the main text with those from MF and DC-MF dynamics under focused sampling, but without a renormalization of the TDM, which we refer to as MF(f) and DC-MF(f), respectively. By presenting these results, we again draw a comparison with full-quantum calculations based on CISD. The applied parameters are identical to those reported in the main text, including the number of trajectories invoked. Transient excited-state occupancies are shown in Figs.~\ref{small} and \ref{long_population} for the short-cavity and long-cavity regimes, respectively, where a lack of TDM renormalization is seen to generally yield an underestimation of the dynamics (such as Rabi oscillations). Included in Fig.~\ref{long_population} are results from MF and DC-MF dynamics under Wigner sampling, obtained by averaging over 5000 and 500 trajectories, respectively. In Fig.~\ref{long_intensity}, we show the optical field intensity profiles for the long-cavity regime, which also includes Wigner sampling results, obtained by averaging over 100\;000 trajectories.



